\documentclass[12pt]{article}

\usepackage{graphicx}
\usepackage{epstopdf}

\def\onehalf{{\textstyle \frac12}}
\def\ii{{\rm i}}
\def\dd{{\rm d}}
\def\rr{{\bf r}}
\def\pp{{\bf p}}
\def\ssr#1{{\scriptscriptstyle\rm #1}}
\def\of#1{{{\scriptstyle(}#1{\scriptstyle)}}}
\def\abs#1{{{\scriptstyle|}#1{\scriptstyle|}}}
\def\parder#1#2{\frac{\partial #1}{\partial #2}}

\def\tsty#1#2{{\textstyle\frac{#1}{#2}}}
\def\matdos#1#2#3#4{\left(\matrix{\displaystyle{#1}&
			\displaystyle{#2}{}_{\mathstrut}
		\cr \displaystyle{#3}&\displaystyle{#4}\cr} \right)}

\def\vecdos#1#2{\left( \matrix{\displaystyle{#1}{}_{\mathstrut}\cr 
		\displaystyle{#2}\cr}\right)}
\def\jour#1#2#3#4{{\it #1{}} {\bf #2}, #3 (#4)}
\def\lab#1{\label{eq:#1}}
\def\rf#1{(\ref{eq:#1})}

\newcommand{\be}{\begin{equation}}
\newcommand{\ee}{\end{equation}}
\newcommand{\bea}{\begin{eqnarray}}
\newcommand{\eea}{\end{eqnarray}}

\begin{document}

\begin{center}

{\Large\bf Superintegrable classical Zernike system} 

\bigskip
	George S.\ Pogosyan,\footnote{Departamento de Matem\'aticas, 
	Centro Universitario de Ciencias Exactas e Ingenier\'ias, 
	Universidad de Guadalajara, M\'exico; Yerevan State University, 
	Yerevan, Armenia; and Joint Institute for Nuclear Research, 
	Dubna, Russian Federation.} Kurt Bernardo Wolf\footnote{Instituto
	de Ciencias F\'isicas, Universiad Nacional Aut\'onoma de M\'exico,
	Cuernavaca.} and Alexander Yakhno\footnote{Departamento de 
	Matem\'aticas, Centro Universitario de Ciencias Exactas e 
	Ingenier\'ias, Universidad de Guadalajara, M\'exico.}
	
\vskip1cm


\vskip1cm

\end{center}

\begin{abstract}
	We consider the differential equation that Zernike proposed
	to classify aberrations of wavefronts in a circular pupil,
	as if it were a classical Hamiltonian with a non-standard potential.
	The trajectories turn out to be closed ellipses. We show that 
	this is	due to the existence of higher-order invariants that 
	close into a cubic Higgs algebra. The Zernike classical system
	thus belongs to the class of superintegrable systems.	
	Its Hamilton-Jacobi action separates in three vertical 
	projections of polar coordinates of a sphere, polar and
	equidistant coordinates on half-hyperboloids, and also in elliptic 
	coordinates on the sphere.
\end{abstract}

\section{Introduction: the Zernike operator}  \label{sec:one}

In Reference \cite[p.\ 700]{Zernike34}, Frits Zernike proposed a 
two-dimensional differential equation whose polynomial solutions 
provide an orthogonal basis for functions $f\of\rr$ in a 
Hilbert space ${\cal L}^2_\ssr{Z}({\cal D}_1)$ over the unit disk 
$\rr\in{\cal D}_1$, $\abs{\rr}\le1$ which ---importantly--- 
have a constant absolute value on the boundary circle: 
$|f(\rr)|_{\abs{\rr}{=}1}=1$.
This Zernike basis is thus distinct from the well-known bases of 
Bessel functions over the disk whose values (or logarithmic
derivatives) vanish on a boundary circle. 
The differential operator and eigenvalue equation of Zernike are
\be 
        \hat Z^{(\alpha,\beta)}f\of\rr := \Big(\nabla^2 + 
                \alpha(\rr\cdot\nabla)^2 + \beta\,\rr\cdot\nabla \Big)
                        f\of\rr = -E\,f\of\rr.
                        \lab{Zernikeq}
\ee
The requirement that this operator be self-adjoint 
under the inner product $(f_1,f_2)_{{\cal D}_1}
:=\int_{{\cal D}_1}\dd^2\rr\,f_1\of\rr^*\,f_2\of\rr$,
i.e., $(\hat Z f_1,f_2)_{{\cal D}_1}
= (f_1,\,\hat Z\,f_2)_{{\cal D}_1}$, constrains the 
coefficients to have the values 
$(\alpha_\ssr{Z},\beta_\ssr{Z}):=(-1,-2)$ \cite{Zernike34}.
In this paper however, we let $\alpha$ and $\beta$ take arbitrary
real values, to be later constrained to those regions
that lead to the closed orbits that we consider to 
be the main feature of interest of the Zernike system.

For $\hat Z^{(\alpha_\ssr{Z},\beta_\ssr{Z})}$ in \rf{Zernikeq}, 
the polar factored solutions $Z_{n,m}(r)\exp(\ii m\phi)$,
$|m|\le n$, correspond to the eigenvalues $E=n(n+2)$; when
normalized to $Z_{n,m}(1)=1$, the radial functions are the 
Zernike polynomials \cite{Zernike34}.
These can be related to the Jacobi polynomials 
$\sim P_n^{(m-n,0)}(2r^2-1)$ whose interval of orthogonality
is $|_{-1}^1\leftrightarrow r|_0^1$. It was remarked in 
Ref.\ \cite{Bhatia-Wolf} that the reasons for postulating 
Eq.\ \rf{Zernikeq} were rather arbitrary, so its authors used the 
Gram-Schmidt method to find the same polynomial solutions from 
first principles. Zernike polynomials have wide applications 
in the correction of optical aberrations by describing 
wavefronts at circular pupils 
(see for example Ref.\ \cite{Born}); they also display 
a host of enticing mathematical properties 
\cite{Myrick,Kintner,Shakibaei,Tango,Wunsche,Ismail}
that are characteristic of algebraic structures.

When $\alpha=0$, $\hat Z^{(0,\beta)}$ reduces to 
a linear combination of generators of the real 
symplectic algebra ${\sf sp(}$4${\sf,R)}$ under Poisson
brackets or commutators \cite[Sect.\ 11.4]{GeomOpt};
when also $\beta=0$, then \rf{Zernikeq} becomes 
simply the Laplace equation with plane wave solutions 
$\sim\exp(\ii{\bf k}\cdot{\rr})$, $|{\bf k}|^2=E$ or, adapted
to polar coordinates $(r,\phi)$, multipole solutions 
$\sim J_m(kr)e^{\ii m\phi}$ with Bessel functions, where
the radial wavenumber $k$ may or may not be quantized
according to whether the boundary conditions are set at a 
finite or infinite radius. 
On the other hand, when $\alpha\neq0$ but $\beta=0$, the
Zernike equation \rf{Zernikeq} reduces to the kinetic
part of a nonlinear oscillator Hamiltonian \cite{Carinena0}.
We shall keep their generic values 
$(\alpha,\beta)\in{\cal R}^2$ and particularize when 
convenient.

We found that it is of interest to examine the {\it classical
counterpart\/} of the Zernike system, which in `wave' (or 
quantum mechanical) form is \rf{Zernikeq}. The process of 
{\it de-quantization\/} of this equation consists in 
replacing
\bea 
        \nabla\mapsto\ii\pp=\ii\vecdos{p_x}{p_y},
        	&& \rr=\vecdos xy,\quad r:=\abs\rr,
                \lab{de-quant1}\\
        \nabla^2\mapsto -(p_x^2+p_y^2)= -\bigg(p_r^2 +
			\frac{p_\phi^2}{r^2}\bigg), 
                &&\rr\cdot\nabla\mapsto\ii(xp_x+yp_y)= \ii\,r p_r.
\lab{de-quant2}
\eea
The operator \rf{Zernikeq} thus yields a classical 
Hamiltonian function $H^{(\alpha,\beta)}=-\hat Z^{(\alpha,\beta)}$ which depends 
on two coordinates and two momenta. 
In Cartesian and polar coordinates, it is 
\bea 
        H^{(\alpha,\beta)}&:=& (p_x^2+p_y^2)
                +\alpha(xp_x+yp_y)^2 -\ii\beta(xp_x+yp_y) \lab{H1}\\
              &=& (1+\alpha r^2)p_r^2 + p_\phi^2/r^2 
                      - \ii\beta rp_r,  \lab{H3}
\eea
and its value is the energy $E$. The appearance of $\ii=\sqrt{-1}$ in 
this Hamiltonian seems indeed anomalous, yet our calculations will 
show that at the end we have a purely real classical system whose 
trajectories can be found explicitly.  

        The Hamilton-Jacobi method is particularly apt to
solve this system, where we shall preferentially use
the polar coordinates $(r,\phi)$ and their momenta $(p_r,p_\phi)$
in \rf{H3}. Since $H^{(\alpha,\beta)}=E$ is independent 
of time and the angular coordinate $\phi$ is cyclic,
the {\it action\/} function $S(r,\phi)$ (also called 
Hamilton's {\it principal\/} function) that satisfies the
Hamilton-Jacobi equation $H+\partial S/\partial t=0$ can be
separated in the form 
\be 
        S(r,\phi) = R\of{r} + p_\phi\phi - Et.
        \lab{act-SR}
\ee
The space derivatives of this function yield the 
polar momenta $p_r$ and $p_\phi$ as
\be 
         p_r=\parder{S(r,\phi)}{r},\quad r=-\parder{S(r,\phi)}{p_r},\qquad 
         p_\phi=\parder{S(r,\phi)}{\phi},\quad
		\phi=-\parder{S(r,\phi)}{p_\phi}.
                                \lab{act-mom}
\ee

In Sect.\ \ref{sec:two} we shall use the derivatives of \rf{act-SR} 
with respect to the radius $r$ and the angle $\phi$,
to find the {\it geometric\/} trajectories $r(\phi)$, 
which are closed ellipses. Then in Sect.\ \ref{sec:three} the 
{\it dynamical\/} trajectories ${\bf r}(t)$ will be found 
differentiating the action $S(r,\phi)$ with respect to the 
energy. The symmetries behind the closure of the orbits will
be elucidated in Sect.\ \ref{sec:four}, where Eq.\ \rf{Zernikeq}
is separated in three spherical, six hyperbolic, and elliptic
coordinates, and shown to lead to constants of motion.
In Sect.\ \ref{sec:five} we show that the operators which
characterize these constants close into a cubic superintegrable
algebra, and offer some additional comments.


\section{Geometric trajectories $r(\phi)$}   \label{sec:two}

The derivative of the action function \rf{act-SR} with
respect to the radius $r$ is the radial momentum,
\be 
        p_r=\parder{S(r,\phi)}{r}=\parder{R\of{r}}{r}.
                        \lab{printotderR}
\ee
Replacing $p_r$ in \rf{H3} yields a quadratic algebraic 
equation for the derivative of $R\of{r}$, namely 
\be 
     (1+\alpha r^2)\bigg(\parder{R\of{r}}{r}\bigg)^2 
                     -\ii\beta r \bigg(\parder{R\of{r}}{r}\bigg)
                     + \frac{p_\phi^2}{r^2} = E,
                                        \lab{eqR1}
\ee
whose two solutions are
\be 
        \parder{R\of{r}}{r} = \frac{\ii\beta\,r 
                \pm \sqrt{-\beta^2r^2 - 4(1+\alpha r^2)(p_\phi^2/r^2 - E)}
                                }{2(1+\alpha r^2)}.  \lab{dFdr}
\ee
From here we find $R\of{r}$ through the indefinite integral
\be
        R\of{r} = \int \dd r \left( \frac{\ii\beta\,r}{2(1+\alpha r^2)}
                \pm \frac{\sqrt{(\alpha E{-}\tsty14\beta^2)r^2+(E{-}\alpha
                   p_\phi^2)-p_\phi^2/r^2}}{1+\alpha r^2}\right).  \lab{Rint}
\ee

        We can now find the trajectories that relate $r$ and $\phi$ by
differentiating \rf{act-SR} with respect to $p_\phi$, 
\be 
        \parder{S(r,\phi)}{p_\phi} = \parder{R\of{r}}{p_\phi} + \phi =\phi_o,
                \lab{parSpphi}
\ee
where $\phi_o$ is a constant of the motion given by
the initial conditions. The derivative of $R\of{r}$
in \rf{Rint} with respect to $p_\phi$, is then
\bea
    \parder{R\of{r}}{p_\phi} &=& \pm\int \dd r \,\parder{}{p_\phi}
         \frac{\sqrt{(\alpha E{-}\tsty14\beta^2)r^2-(\alpha+1/r^2)p_\phi^2+E}
                        }{1+\alpha r^2}  \lab{int-diff1}\\
         &=&\mp p_\phi\int \frac{\dd r}{r} \,\frac{1}{
                \sqrt{(\alpha E{-}\tsty14\beta^2)r^4 
                + (E-\alpha p_\phi^2)r^2 - p_\phi^2}}
                                         \lab{int-difff2}\\
        &=&\mp \frac{p_\phi}{2} \int \dd z \,\frac{1}{
                z \sqrt{a + b\,z + c\,z^2}} \lab{int-diff3}
\eea
where in the last equality we have substituted $z=r^2$ with 
$\dd r/r = \onehalf \dd z/z$, and we define 
\be
        a:=-p_\phi^2, \quad b:= E-\alpha p_\phi^2, \quad 
                 c:=\alpha E-\tsty14\beta^2. \lab{uvw}
\ee
We note that the imaginary summand in \rf{Rint} is absent 
from this equation and thus from the system. The double sign in
\rf{int-diff1} corresponds to the $\pm p_\phi$ angular momentum
of a trajectory traversed in opposite directions. 

One finds the indefinite integral solved in \cite[Eqs.\ 2.266]{GR},
with various expressions involving inverse trigonometric and
hyperbolic functions, or logarithms, depending on the signs of 
the constants; in our case \rf{uvw} $a<0$ and for $b^2-4ac
=(E+\alpha p_\phi^2)^2 - \beta^2p_\phi^2>0$, the integral is 
\be 
        \int \dd z \,\frac{1}{z \sqrt{a + b\,z + c\,z^2}}
                =\frac1{\sqrt{-a}}\arcsin \frac{2a+bz}{z\sqrt{b^2-4ac}}.
                                \lab{Int266}
\ee
Thus, joining Eqs.\ \rf{parSpphi}, \rf{uvw}, and \rf{Int266}, 
we obtain
\be
        \phi-\phi_o = -\parder{R\of{r}}{p_\phi} = \frac12\arcsin
                \frac{(E-\alpha p_\phi^2)r^2 -2p_\phi^2
                        }{r^2\sqrt{(E+\alpha p_\phi^2)^2 - \beta^2p_\phi^2}},
                                \lab{phiphio}
\ee
and this leads to $\phi(r^2)$ in the form
\be 
        \sin2(\phi{-}\phi_o)=\frac{Ar^2-B}{Cr^2},\quad \left\{
                \begin{array}{l} A:=E-\alpha p_\phi^2,\\
                        B:=2p_\phi^2,\\
                        C:=\sqrt{(E+\alpha p_\phi^2)^2-\beta^2p_\phi^2}.
                        \end{array}\right.  \lab{sindosphi}
\ee

\begin{figure}[t]
\centering  
\centerline{\includegraphics[width=1.0\columnwidth]{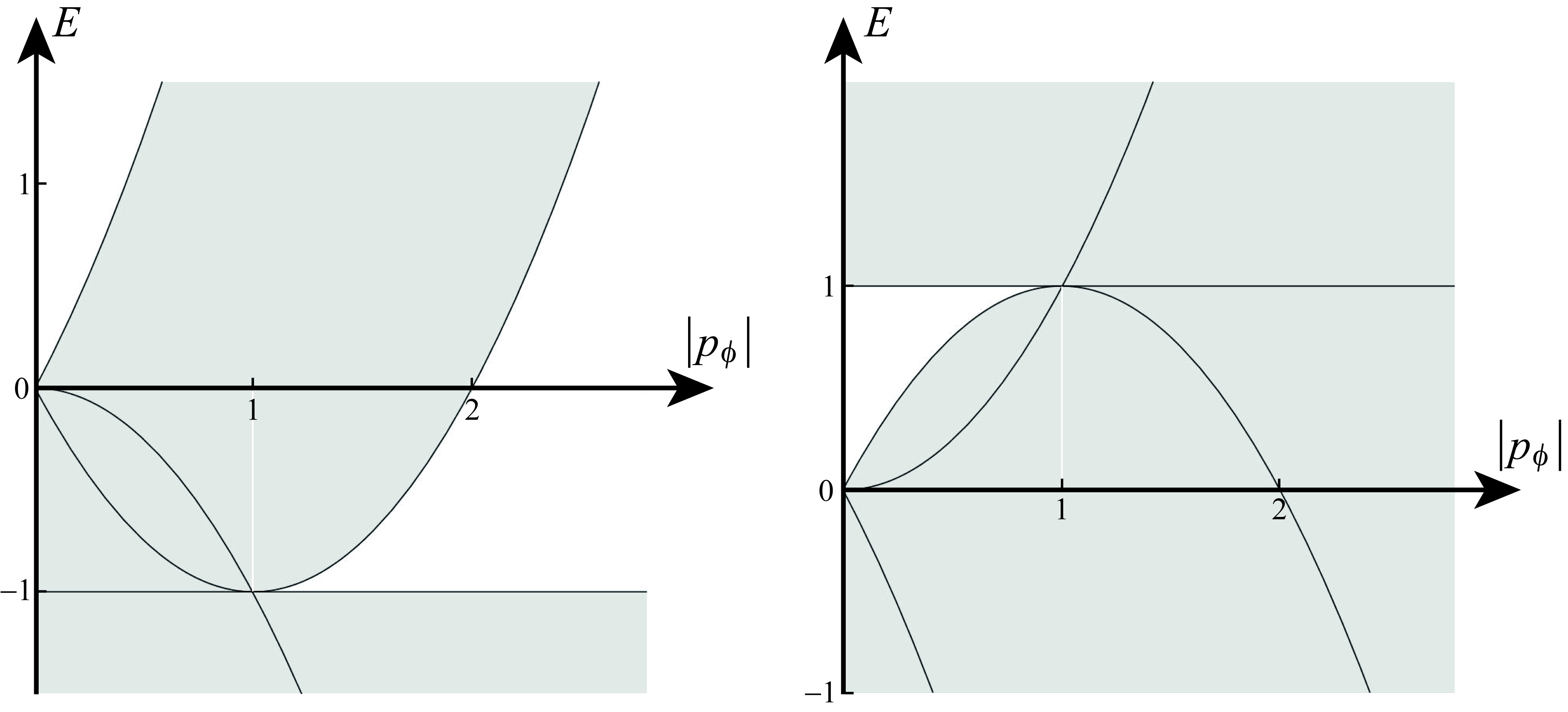}}
\caption[]{Regions of the plane of angular momentum 
$p_\phi$ and energy $E$, where closed trajectories 
are allowed by the inequalities \rf{cond1} (in white) for 
$(\alpha,\beta)$-Zernike systems. {\it Left\/}: Allowed
regions for the original Zernike system  
$(\alpha_\ssr{Z},\beta_\ssr{Z})=(-1,-2)$.  {\it Right\/}: 
Allowed regions for $(\alpha,\beta)=(1,2)$.  Closed elliptical 
trajectories do not occur in the gray regions.
The structure of these regions is generic for all $\alpha$ and 
$\beta\neq0$. The units of $E$ in these graphs are 
$\beta^2/4|\alpha|$ and the units of $|p_\phi|$ 
are $\frac12|\beta/\alpha|$.} 
\label{fig:Epregion}
\end{figure}

We can invert the dependence to $r(\phi)$ by solving for the square 
radius and setting for convenience $\phi_o=-\frac14\pi$, 
\bea 
        r^2(\phi)&=&\frac{B}{A-C\cos2\phi}
                = \frac{2p_\phi^2}{\displaystyle (E{-}\alpha p_\phi^2)
                        -\sqrt{(E{+}\alpha
		p_\phi^2)^2-\beta^2p_\phi^2}\,\cos2\phi}    \lab{rcuadrada}\\
                &=& \frac{D}{1-\varepsilon\cos2\phi},\quad\left\{
                                \begin{array}{l}
               D:=B/A=2p_\phi^2/(E{-}\alpha p_\phi^2),
               	\quad E\neq\alpha p_\phi^2,\\[3pt] 
          \displaystyle \,\varepsilon\,:=\frac{C}{A}  
          	=\frac{\sqrt{(E+\alpha p_\phi^2)^2-\beta^2p_\phi^2}}{E-\alpha p_\phi^2}.
                       \end{array}\right. \lab{rrcuadrada}
\eea
This is the parametric equation for ellipses, provided that
\be 
        \begin{array}{lclcl} 
        \varepsilon\hbox{ real}& \Rightarrow & C^2\ge0&
        \Rightarrow& \displaystyle 
                \left\{\begin{array}{l} E\le -\alpha p_\phi^2-|\beta p_\phi|,\\
                 E\ge -\alpha p_\phi^2+|\beta p_\phi|,\end{array}\right.\\
        \abs\varepsilon<1& \Rightarrow & A^2>C^2 & 
          \Rightarrow &\quad\, 4\alpha E<\beta^2,\\
         r^2(\phi)>0& \Rightarrow &D>0&\Rightarrow &\quad\,  E>\alpha p_\phi^2.
         \end{array}
                \lab{cond1}
\ee
These conditions restrict the range of energies $E$ and
angular momenta $p_\phi$ where the trajectories are
real and closed. As shown in Fig.\ \ref{fig:Epregion} (left)
for the generic Zernike range $\alpha<0$, $\beta\neq0$, 
the first condition {\it excludes\/} the energy interval 
between the two parabolas, $-\alpha\,p_\phi^2-|\beta\, p_\phi|
\le E \le -\alpha\,p_\phi^2+|\beta\,p_\phi|$;
the second inequality is (for $\alpha<0$) a {\it lower bound\/} 
$E>-\beta^2/4|\alpha|$ (equal to $-1$ for the Zernike 
case); lastly, the third condition excludes the
interior of the parabola $E=\alpha p_\phi^2$ that
has its apex at the origin, and which eliminates the
region $|p_\phi|<-\onehalf |\beta|/\alpha$ that was 
left allowed by the previous two conditions. 

In Fig.\ \ref{fig:Epregion} (right) we show
the allowed regions for the generic Zernike range 
$\alpha>0$, $\beta\neq0$. The two parabolas 
stemming from the first inequality in \rf{cond1},
under $\alpha\leftrightarrow -\alpha$ reflect the
$E$-axis; the second inequality in \rf{cond1} is 
now the {\it upper\/} bound $E<\beta^2/4\alpha$; and 
the third inequality allows elliptic orbits 
in the remaining interior of the parabola, namely 
$-\alpha p_\phi^2+|\beta p_\phi| <E<\beta^2/4\alpha$ for 
$0\le|p_\phi|<|\beta|/2\alpha$. 
Finally, when $\alpha=0$, the 
`forbidden' region between the two
parabolas due to the first condition in 
\rf{cond1} becomes $-|\beta p_\phi|\le E \le |\beta p_\phi|$,
while the second two conditions are satisfied by $E>0$,
so that closed elliptical trajectories occur for all 
$E\ge|\beta p_\phi|$ .

Since we took 
$\phi_o=-\frac14\pi$, the $y$-axis is at 
$\phi=0$ and the $x$-axis at $\phi=\frac12\pi$.  
The semi-major and semi-minor axes of the ellipse are, 
respectively,
\be 
                \mu_y:=\sqrt{\frac{D}{1-\varepsilon}}
                        =\sqrt{\frac{B}{A-C}},\quad
                \mu_x:=\sqrt{\frac{D}{1+\varepsilon}}
                        =\sqrt{\frac{B}{A+C}}. \lab{axes}
\ee
The {\it area\/} of this ellipse is given by $\pi$ times the product
of the two semi-axes,
\be 
        \hbox{area}=\pi \mu_x\mu_y=\frac{\pi D}{\sqrt{1{-}\varepsilon^2}}
                =\frac{\pi B}{\sqrt{A^2{-}C^2}}
                = \frac{2\pi\, |p_\phi|}{\sqrt{\beta^2-4\alpha E}}.
                        \lab{area-ellipse}
\ee


\section{Dynamical trajectories $r(t)$ and orbits}   \label{sec:three}

We return now to the integral expression for $R\of{r}$ in \rf{Rint},
differentiating the action $S(r,\phi)$ in \rf{act-SR} now with respect 
to the energy $E$,
\be 
        \parder{S(r,\phi)}{E} = \parder{R\of{r}}{E} - t =-t_o,
                \lab{parSE}
\ee
where $t_o$ is the initial time constant. Instead of 
\rf{int-diff1}--\rf{int-diff3}, we now have 
\bea
        \parder{R\of{r}}{E} &=& \pm\int \dd r \,\parder{}{E}
                \frac{\sqrt{(\alpha E{-}\tsty14\beta^2)r^2-(\alpha+1/r^2)p_\phi^2+E}
                        }{1+\alpha r^2}  \lab{Eint-diff1}\\
        &=&\pm \frac12\int \dd r \,\frac{1}{
                 \sqrt{(\alpha E{-}\tsty14\beta^2)r^2-(\alpha+1/r^2)p_\phi^2+E}}
                         \lab{Eint-diff2}\\                         
        &=&\pm \frac14 \int \dd z \,\frac{1}{
                 \sqrt{a + b\,z + c\,z^2}} \lab{Eint-diff4}
\eea
where as before we have set $z=r^2$, and $a,\,b,\,c$ are again
given by \rf{uvw}. The indefinite integral can be found in
\cite[Eqs.\ 2.261]{GR}; it is
\be 
        \int \dd z \,\frac{1}{\sqrt{a + b\,z + c\,z^2}}
                =\frac{-1}{\sqrt{-c}}\arcsin \frac{2cz+b}{\sqrt{b^2-4ac}},
                                \lab{Int261}
\ee
The conditions for this integral to be proper, $c<0$
and $b^2-4ac>0$ also lead to \rf{cond1}, while the solutions
corresponding to \rf{sindosphi} are now 
\be 
        \sin\Big(4(t{-}t_o)\sqrt{U}\Big)=\frac{A-2U\,r(t)^2}{C},\quad 
           U:=\tsty14\beta^2 -\alpha E  = \frac{A^2-C^2}{2B} >0,
                         \lab{sindost}
\ee
with $A$ and $C$ given by \rf{sindosphi}.

From here we can extract the dependence of the square 
radius of the trajectory on time as \rf{rrcuadrada} did
for the angle. We choose $t_o$ such that $r(t)|_{t=0}=\mu_y$ is 
the semi-major axis in \rf{axes}, i.e., 
$4t_o\surd U=\onehalf\pi$, so $t_o=\frac18\pi/\surd U$, 
and write 
\be 
	\begin{array}{rcl}       
        r^2(t)&=&\displaystyle \frac{A + C \cos(4t\sqrt{U})}{2U}\\
        	&=& \displaystyle \frac{E+\displaystyle\sqrt{(E{+}\alpha p_\phi^2)^2
        	{-}\beta^2p_\phi^2}\,\cos(2t\sqrt{\beta^2{-}4\alpha E})
              -\alpha p_\phi^2}{\tsty12\beta^2-2\alpha E}. 
                          \end{array}               \lab{trcuadrada}
\ee        
This is a periodic function of time, with period $4T\surd U=2\pi$, or
\be 
        T=\pi\Big/ \sqrt{\beta^2-4\alpha\,E}. \lab{period}
\ee
In the generalized Zernike range $\alpha<0$, the radicand
is positive; when $\alpha>0$, the second inequality in 
\rf{cond1} prevents the orbits from being closed for 
$\alpha E>\frac14\beta^2$.
Although orbits in the Zernike range are ellipses, 
they {\it differ\/} from the isochronous orbits of the 
classical harmonic oscillator, whose period does not depend 
on their energy \cite{Carinena}. 

\begin{figure}[t]
\centering  
\centerline{\includegraphics[width=0.6\columnwidth]{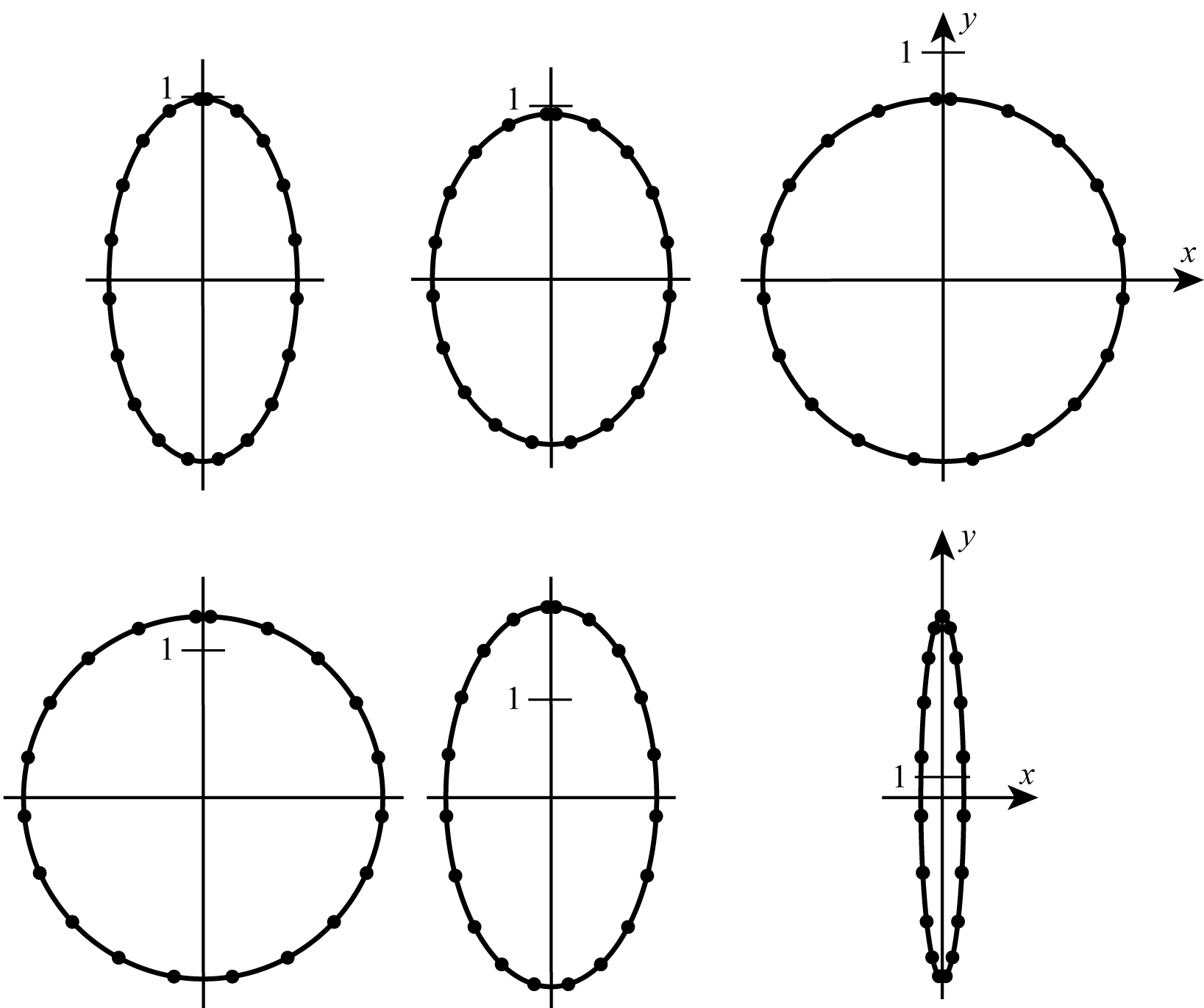}}
\caption[]{Trajectories $\Big(x(t),y(t)\Big)$ in the
classical Zernike system $(\alpha_\ssr{Z},\beta_\ssr{Z})=(-1,-2)$
and angular momentum $p_\phi=3$, for equidistant times 
$t\in[0, T]$. {\it Upper row\/}: trajectories inside the disk
${\cal D}_1$ for energies $E=35$, 20, and 15 (at the lower 
boundary of the upper allowed region of Fig.\ \ref{fig:Epregion}). 
{\it Lower row\/}: trajectories {\it outside\/} the 
unit disk ${\cal D}_1$, for energies $E=3$ (at the upper boundary of the
lower allowed region), 1, and $-0.9$ (near to the lower forbidden 
region), which fall completely outside the disk and correspond to
the hyperbolic case to be seen in Sect.\ \ref{sec:four}. 
We mark the scale 1 on the $y$-axis, understood to be
in units of $1/\surd|\alpha|$.} 
\label{fig:trajectories}
\end{figure}

As a function of time, the trajectories $\Big(x(t),\,y(t)\Big)$ can be found 
from the previous expressions, \rf{rrcuadrada} and \rf{trcuadrada}, as
\bea         
      x(t)&=&r\sin\phi=r\sqrt{\onehalf(1{-}\cos2\phi)}
            =\frac1{\sqrt{2\varepsilon}}\sqrt{(\varepsilon{-}1)\,r^2(t)+D}
                        \nonumber\\ &=&                         
        \frac1{\sqrt{2\varepsilon}}\sqrt{(\varepsilon{-}1) \frac{A + C \cos(4t\sqrt{U})}{ 2U} + D},
                                       \lab{xoft}\\
      y(t)&=&r\cos\phi=r\sqrt{\onehalf(1{+}\cos2\phi)}
                =\frac1{\sqrt{2\varepsilon}}\sqrt{(\varepsilon{+}1)\,r^2(t)-D}
                        \nonumber\\ &=&                         
           \frac1{\sqrt{2\varepsilon}}\sqrt{(\varepsilon{+}1)\frac{A + C \cos(4t\sqrt{U})}{ 2U} - D},
                                                \lab{yoft}
\eea
and are shown in Fig.\ \ref{fig:trajectories} for the 
Zernike case $(\alpha_\ssr{Z},\beta_\ssr{Z})=(-1,-2)$,
but are valid for the range $\alpha<0$.

The trajectories are circular 
when $\varepsilon=0$, i.e., $C=0$ or $E+\alpha p_\phi^2
=\pm|\beta\,p_\phi|$. This is the case of the upper 
right and lower left trajectories in Fig.\ 
\ref{fig:trajectories}. For $\alpha<0$ it occurs 
on the two parabolas that bound the region excluded by the 
first condition in \rf{cond1} and respect the other two
inequalities.  The radius of
those circles can be found from \rf{rrcuadrada},
as $r^2(\phi)=D$. At the upper boundary one has  
$E=-\alpha p_\phi^2 +|\beta\,p_\phi|\ge -\alpha p_\phi^2$,
so in the Zernike $\alpha<0$ region this means 
$E\ge |\alpha|p_\phi^2$, which in turn entails that 
$|\alpha|B\le A$, or $D\le1/|\alpha|$, which yields the
radius of the circle as $r_\circ=1/\surd|\alpha|$; in
the case $\alpha_\ssr{Z}=-1$ this is the boundary of
the unit circle of Zernike's differential equation \cite{Zernike34}. 
On the other hand, at the lower boundary in the same Zernike
range $\alpha<0$, $E=|\alpha| p_\phi^2 -|\beta\,p_\phi|$, and one has
$r^{\prime\,2}_\circ=D=2p_\phi^2/(2|\alpha|p_\phi^2-|\beta p_\phi|)>1/|\alpha|$, 
which for $\alpha_\ssr{Z}=-1$ exceeds the unit radius allotted 
by Zernike's requirement. We conclude that the elliptic  
trajectories in the lower `allowed' region of Fig.\ 
\ref{fig:Epregion} (left) cannot correspond with solutions of 
the Zernike differential equation \rf{Zernikeq}. Only those
in the upper region do. On the other extreme of the $\alpha<0$ 
region, the trajectories become lines when
$\varepsilon\to1$, namely for ever larger $E$  
and also when $E$ approaches the lower boundary 
$-\beta^2/4|\alpha|$.

Regarding the region $\alpha>0$ in Fig.\ \ref{fig:Epregion} 
(right), the excentricity  in \rf{rrcuadrada} 
is $\varepsilon=0$ on the parabola $E=-\alpha p_\phi^2 +|\beta\,p_\phi|$.
The radii of those circles can be found as we did above, yielding 
$r_\circ^2(\phi)=2p_\phi^2/(|\beta p_\phi|-2\alpha p_\phi^2)$.
The trajectory is a {\it unit\/} circle when 
$2(1+\alpha)p_\phi^2=|\beta p_\phi|$, i.e., 
$|p_\phi|=|\beta|/2(\alpha{+}1)<|\beta|/2\alpha$. 
This value falls on a single point of the parabolic boundary 
of the allowed region in Fig.\ \ref{fig:Epregion} (right).
On the upper boundary of that region, $E=\beta^2/4\alpha$, 
the excentricty is $\varepsilon=1$ and the trajectores are
lines.  Finally, when $\alpha=0$ and the allowed region is 
$E\ge|\beta p_\phi|$, on its boundary we have $\varepsilon=0$
circles of radii $r_\circ^2=2|p_\phi/\beta|$.


\section{Separation of variables and symmetries}   \label{sec:four}

        The classical Zernike Hamiltonian \rf{H1} in Cartesian 
coordinates can be subject to the Hamilton-Jacobi method of 
solution with the action partial derivatives 
$p_x=\partial S/\partial x$ and $p_y=\partial S/\partial y$, and
yields the Hamiltonian \rf{H1} written as
\be
        H = \left(\parder Sx\right)^2 +  \left(\parder Sy\right)^2 
        + \alpha\left(x \parder Sx + y \parder Sy\right)^2 
        -\ii\beta \left(x \parder Sx + y \parder Sy\right)= E.  \lab{HJ-1}
\ee 
This equation is separable on the $(x, y)$-plane, but the boundary
condition imposed by Zernike \cite{Zernike34} on the solutions, namely
that their absolute value at the boundary $x^2 + y^2 = 1$ be constant,
can only be separated in polar coordinates, as we did in 
Sect.\ \ref{sec:two}. Although the classical Zernike system 
appears to belong to the class of Bertrand systems \cite{Bertrand} 
in which all bounded orbits are closed, it does not qualify as
such because the linear and quadratic $\rr\cdot\nabla$ terms 
replace the two-dimensional central force potentials of the 
Coulomb or isotropic oscillator systems. 
We surmise that this feature is a specific consequence 
of the superintegrability of the Zernike system. 
It is therefore of interest to find any additional separable 
systems of orthogonal coordinates and, associated with these, 
the extra symmetry operators that will clearly demonstrate 
the classical Zernike Hamiltonian to be superintegrable. 
We remind the reader that in an $N$-dimensional space 
with constant curvature (real or complex), a
maximally superintegrable system allows, in addition to
the Hamiltonian $H$, another $2N-2$ functionally independent 
constants of motion, $L_1$, $L_2$, \dots\,, $L_{2N-2}$,
$L_{2N-1}:=H$, that are in involution with $H$, namely 
$\{H,L_i\}=0$ for $i\in\{1,2,\ldots,2N{-}2\}$ 
\cite{Miller-etal}.

\subsection{Coordinate systems on sphere and hyperboloid}

Equation \rf{Zernikeq} is linear and of second order,
\be
	(1+\alpha x^2) \frac{\partial^2 f}{\partial x^2} 
	+ 2\alpha xy  \frac{\partial^2 f}{\partial x \partial y}  
	+ (1+\alpha y^2) \frac{\partial^2 f}{\partial y^2} 
	+(\alpha+ \beta)\left(x \frac{\partial f}{\partial x} 
	+ y \frac{\partial f}{\partial y}\right) = -E f. \lab{ell-hyp}
\ee
According to the standard classification, this equation is 
of elliptic type when $-\alpha r^2 <1$, of parabolic type 
when $-\alpha r^2 = 1$, and of hyperbolic type 
when $-\alpha r^2 > 1$. The original Zernike case 
$\alpha_\ssr{Z}=-1$ is in the range $\alpha <0$, where 
the region of ellipticity is the interior of the circle 
$r < 1/\surd|\alpha|$. On the other hand, when $\alpha\ge0$, 
the equation \rf{Zernikeq} is of elliptic type over the 
whole $x$-$y$ plane ${\cal R}^2$.

To be within the Zernike case we consider first the range $\alpha<0$, 
and map the open disk $x^2+y^2<1/|\alpha|=:R^2$  on the  
hemisphere $\xi_1^2+ \xi_2^2+\xi_3^2= R^2$, $\xi_3 \ge 0$, 
embedded in a Euclidean space with three Cartesian coordinates 
$\xi_i$, through the orthogonal (or `{\it vertical\/}') projection 
\be
        \xi_1 = x, \quad \xi_2 = y, \quad  
                \xi_3 = \sqrt{R^2-x^2-y^2}.\lab{ort-1} 
\ee
In these coordinates the Hamiltonian 
equation \rf{HJ-1} can be separated 
into three mutually orthogonal spherical systems 
of coordinates \cite{PSW1},
\bea
        &&\hskip-30pt\hbox{System I:} \lab{Syst-I} \\
                &&\hskip-30pt\xi_1 = R \sin\vartheta\cos\varphi,\quad 
                \xi_2= R \sin\vartheta\sin\varphi,\quad
                \xi_3= R \cos\vartheta,\quad
                \vartheta|_0^{\pi/2},\ \varphi|_0^{2\pi},
                \nonumber\\
        &&\hskip-30pt\hbox{System II:} \lab{Syst-II} \\
                &&\hskip-30pt\xi_1 = R \cos\vartheta,\quad 
                \xi_2= R \sin\vartheta\cos\varphi,\quad
                \xi_3= R \sin\vartheta\sin\varphi,\quad
                \vartheta|_0^\pi,\ \varphi|_0^\pi,
                \nonumber\\
        &&\hskip-30pt\hbox{System III:} \lab{Syst-III} \\
                &&\hskip-30pt\xi_1 = R \sin\vartheta\sin\varphi,\quad 
                \xi_2= R \cos\vartheta,\quad
                \xi_3= R \sin\vartheta\cos\varphi,\quad
                \vartheta|_0^\pi,\ \varphi|_{-\frac12\pi}^{\frac12\pi},
                \nonumber
\eea
and in the {\it elliptical\/} system of coordinates \cite{PSW1,LS1,L1} 
to be seen below.  

	Still within the $\alpha<0$ case, we can consider the {\it outside\/}
of the circle at radii $r^2>1/|\alpha|$, where the equation \rf{ell-hyp}
is hyperbolic. There one can map the trajectories of the $x$-$y$ plane
on trajectories on the one-sheeted half-hyperboloid  
$\xi_1^2+\xi_2^2-\xi_3^2=R^2=1/|\alpha|$. Coordinates that permit 
separation of variables for \rf{ell-hyp} replace trigonometric 
functions by hyperbolic functions thus:
\bea
        &&\hskip-30pt\hbox{System H$^\prime$I (pseudo-spherical):} \lab{SystH-I} \\
                &&\hskip-30pt\xi_1 = R \cosh\tau\cos\varphi,\quad 
                \xi_2= R \cosh\tau\sin\varphi,\quad
                \xi_3= R \sinh\tau,\quad
                \tau\in{\cal R},\ \varphi|_0^{2\pi},
                \nonumber\\
        &&\hskip-30pt\hbox{System H$^\prime$II (equidistant):} \lab{SystH-II} \\
                &&\hskip-30pt\xi_1 = \pm R \cosh\tau_1,\quad 
                \xi_2= R \sinh\tau_1\sinh\tau_2,\quad
                \xi_3= R \sinh\tau_1\cosh\tau_2,\quad
                \!\tau_1,\tau_2{\in}{\cal R},\!\!\!{}
                \nonumber\\
        &&\hskip-30pt\hbox{System H$^\prime$III (equidistant):} \lab{SystH-III} \\
                &&\hskip-30pt\xi_1 = R \cosh\tau\sin\varphi,\quad 
                \xi_2= R \cos\varphi,\quad
                \xi_3= R \sinh\tau\sin\varphi,\quad
                \tau\in{\cal R},\ \varphi|_0^{2\pi}.
                \nonumber
\eea

On the other hand when $\alpha > 0$, the region of ellipticity being 
the whole plane ${\cal R}^2$, allows one to map this plane on the upper 
sheet of the two-sheeted hyperboloid 
$\xi_3^2 - \xi_1^2 - \xi_2^2 = \varrho^2 = 1/\alpha$ 
using `modified' coordinate systems:
\bea
        &&\hskip-30pt\hbox{System HI (pseudo-spherical):} \lab{SystHp-I} \\
                &&\hskip-30pt\xi_1 = \varrho \sinh\tau \cos\varphi,\quad 
                \xi_2= \varrho \sinh\tau\sin\varphi,\quad
                \xi_3= \varrho \cosh\tau,\quad
                \tau \in{\cal R},\ \varphi|_0^{2\pi},
                \nonumber\\
        &&\hskip-30pt\hbox{System HII (equidistant):} \lab{SystHp-II} \\
                &&\hskip-30pt\xi_1 = \varrho \sinh\tau_1,\quad 
                \xi_2= \varrho \cosh\tau_1\sinh\tau_2,\quad
                \xi_3= \varrho \cosh\tau_1\cosh\tau_2,\quad
                \tau_1, \tau_2 \in{\cal R},
                \nonumber\\
        &&\hskip-30pt\hbox{System HIII (equidistant):} \lab{SystHp-III} \\
                &&\hskip-30pt\xi_1 = \varrho \cosh\tau'_1 \sinh\tau'_2,\quad 
                \xi_2= \varrho \sinh\tau'_1,\quad
                \xi_3= \varrho \cosh\tau'_1\cosh\tau'_2,\quad
                 \tau'_1, \tau'_2 \in{\cal R}.
                \nonumber
\eea
The hyperboloidal coordinates in \rf{SystH-I}--\rf{SystHp-III} have
been defined in Ref.\ \cite{Pogosyan-Yakhno}.


\subsection{Separation in spherical systems I, H$^\prime$I and HI} 

\begin{figure}[t]
\centering  
\centerline{\includegraphics[width=0.5\columnwidth]{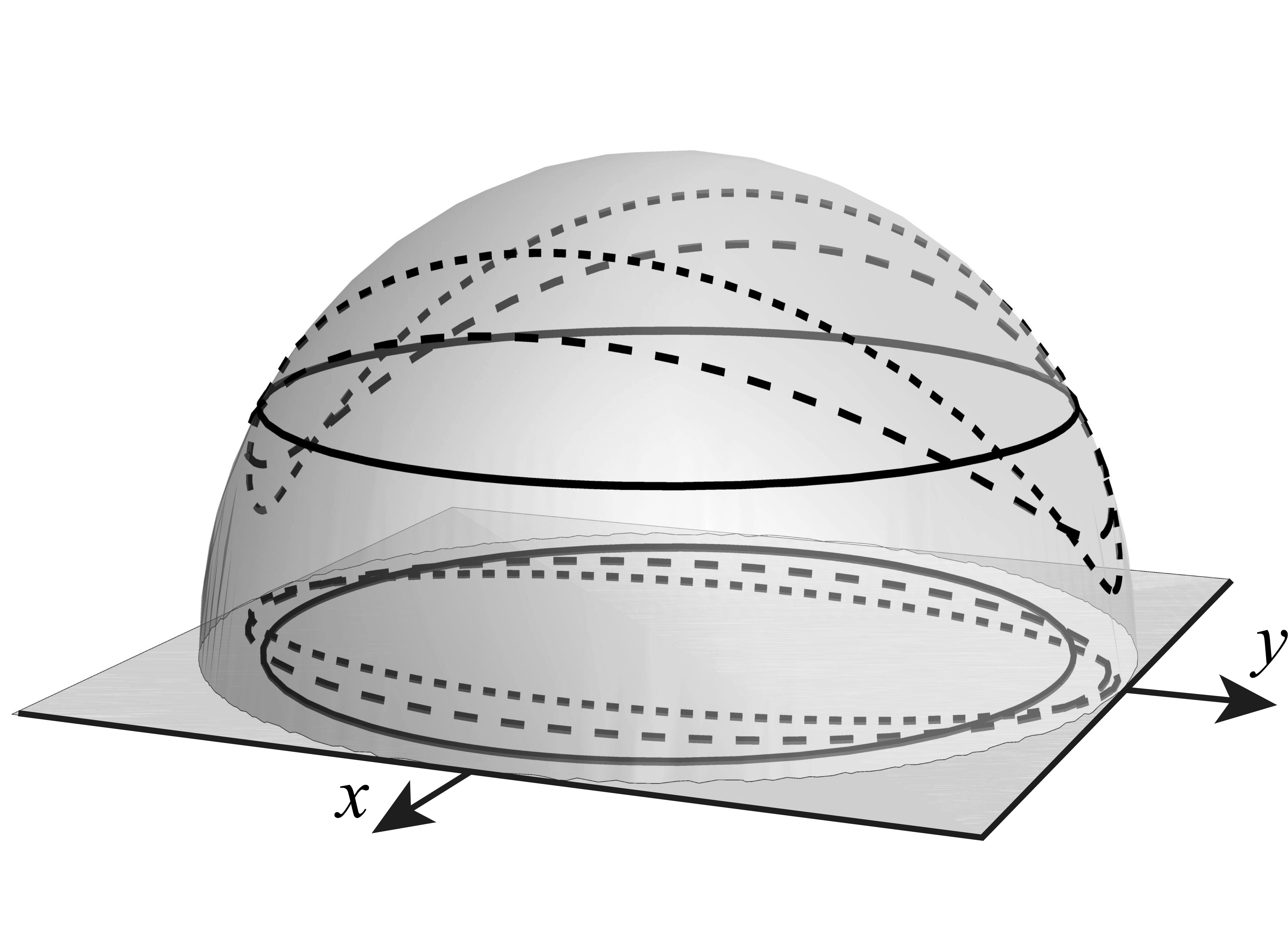}}
\caption[]{Trajectories $\vartheta(\varphi)$ on the hemisphere
given by \rf{HJ-4} for the allowed upper regions of the Zernike 
system $(\alpha_\ssr{Z},\beta_\ssr{Z})=(-1,-2)$ in Fig.\ 
\ref{fig:Epregion} (left), and their projection on the $x$--$y$
plane inside the unit disk ${\cal D}_1$, for the values corresponding 
to the upper row of orbits in Fig.\ \ref{fig:trajectories}: 
$p_\phi=3$ and energies $E=15$ (continuous line, the circular 
orbit at the boundary of the allowed region); $E=20$ (dashed line),
and $E=35$ (dotted line).} 
\label{fig:trajectories_sph}
\end{figure}

In the spherical coordinates ($\vartheta,\varphi$) of System I in 
\rf{Syst-I} for $\alpha<0$, the Hamilton-Jacobi expression in 
\rf{HJ-1} acquires 
the form 
\be
        \frac{1+\alpha R^2\sin^2\vartheta}{R^2 \cos^2\vartheta}
        	\left(\frac{\partial S}{\partial \vartheta}\right)^2 
        - \ii \beta \tan\vartheta\left(\frac{\partial S}{\partial \vartheta}\right)
        +  \frac{1}{R^2 \sin^2\vartheta}
        	\left(\frac{\partial S}{\partial\varphi}\right)^2 
         = E. \lab{HJ-2}
\ee  
This equation is integrable with the help of the first-order 
integral of motion 
\be 
        I_1 := p_\varphi = x p_y - y p_x, 
                \lab{I1}
\ee
that is independent of $(\alpha,\beta)$ and separates the action function as 
$S(\vartheta, \varphi) = S_1(\vartheta) + p_\varphi \varphi$, 
leading to the equation 
\be
      \frac{1+\alpha R^2 \sin^2\vartheta}{R^2 \cos^2\vartheta}  
      	\left(\frac{\dd S_1}{\dd \vartheta}\right)^2 
        - \ii \beta \tan\vartheta\left(\frac{\dd S_1}{\dd \vartheta}\right)
        +  \frac{p^2_\varphi}{R^2 \sin^2\vartheta}  = E. \lab{HJ-3}
\ee    
Using the same approach of Sect.\ \ref{sec:three} for the Zernike 
$\alpha<0$ case, one finds the trajectory $\vartheta(\varphi)$ to be
\be
        \sin^2\vartheta =  \frac{|\alpha|\, D}{1-\varepsilon \cos2\varphi}, 
        	\lab{HJ-4}
\ee 
where $D$ and $\varepsilon$ are given in \rf{rrcuadrada},
and which lies within the hemisphere of radius $R = 1/\surd|\alpha|$, 
as seen in Fig.\ \ref{fig:trajectories_sph}. The trajectories 
reach the rim $\vartheta=\onehalf\pi$ only when $\beta p_\phi=0$.

\begin{figure}[t]
\centering  
\centerline{\includegraphics[width=0.5\columnwidth]{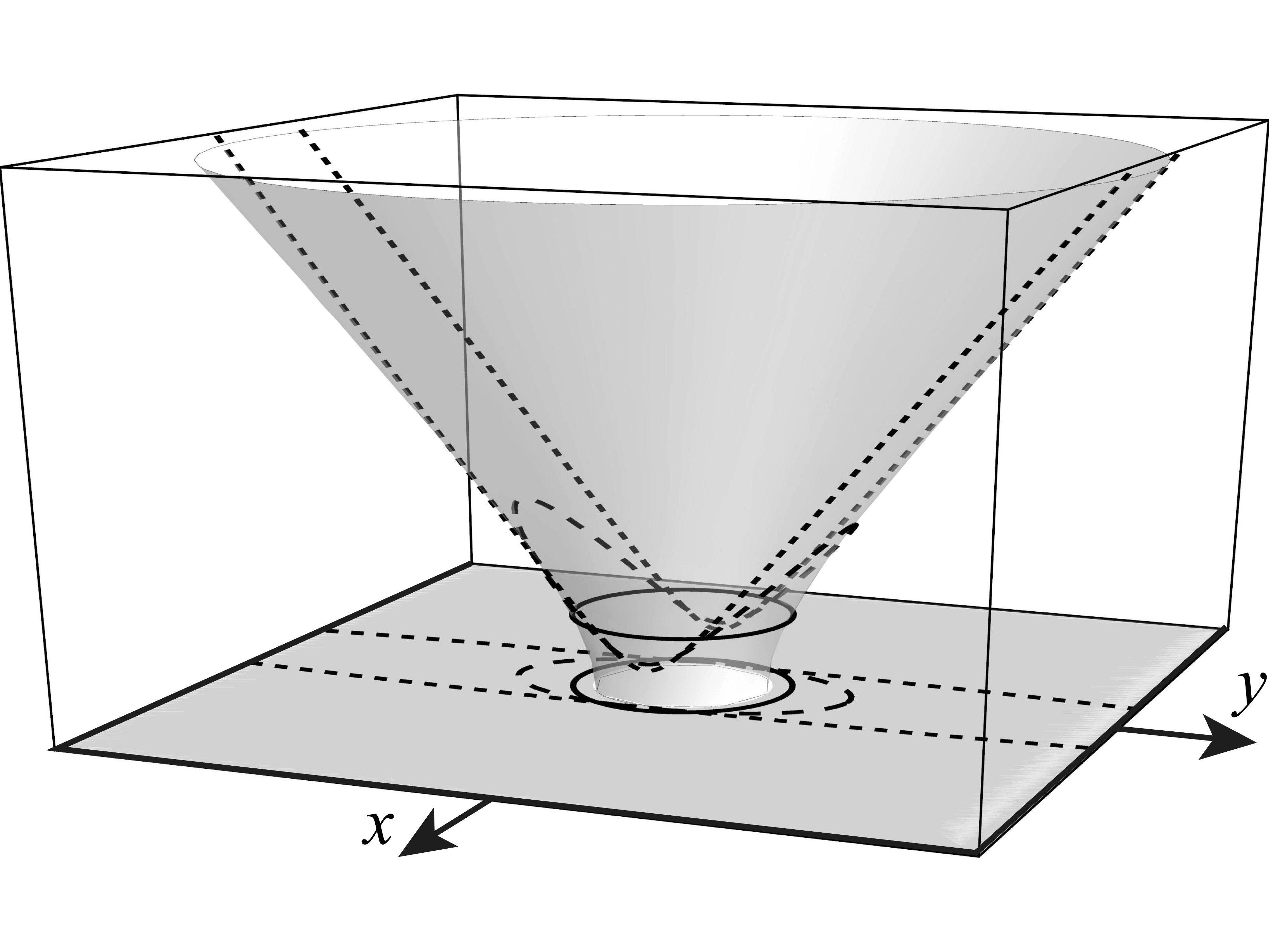}}
\caption[]{Trajectories $\tau(\varphi)$ on the half-hyperboloid
(of one sheet) given by \rf{HJprime-4}, with $\alpha=-1$ and $\beta=-2$,
and their projection on the $x$--$y$ plane outside the unit disk ${\cal D}_1$. 
The parameter values 
are the same as in the second row of Fig.\ \ref{fig:trajectories}, 
namely $p_\phi=3$ and energies $E=3$ (at the upper boundary of the
lower allowed region, marked by a continuous line), 
1 (dashed line), and $-0.9$ (near to the lower forbidden 
region, dotted line).} 
\label{fig:trajectories_hyp1}
\end{figure}

Still in the $\alpha<0$ case, the pseudo-spherical coordinates 
($\tau,\varphi$) of System H$^\prime$I in \rf{SystH-I} allow
separation of the action function as 
$S(\tau, \varphi) = S_1(\tau) + p_\varphi \, \varphi$,  
so the Hamiltonian \rf{HJ-1} leads to the equation
\be
      \alpha\left(\frac{\dd S_1}{\dd \tau}\right)^2 
        - \ii \beta \coth\tau\,\left(\frac{\dd S_1}{\dd \tau}\right)
        -  \frac{\alpha\,p^2_\varphi}{\cosh^2\tau}  = E. 
        \lab{HJprim-3}
\ee 
Then the trajectories, instead of \rf{HJ-4}, are given by
\be
        \cosh^2\tau =  \frac{|\alpha|\, D}{1-\varepsilon \cos2\varphi}, 
        	\lab{HJprime-4}
\ee
with $D$ and $\varepsilon$ given in \rf{rrcuadrada}.
These are closed orbits in the region $r^2>1/|\alpha|$. In Figure
\ref{fig:trajectories_hyp1} we show such trajectories on the 
one-sheeted half-hyperboloid.

\begin{figure}[t]
\centering  
\centerline{\includegraphics[width=0.5\columnwidth]{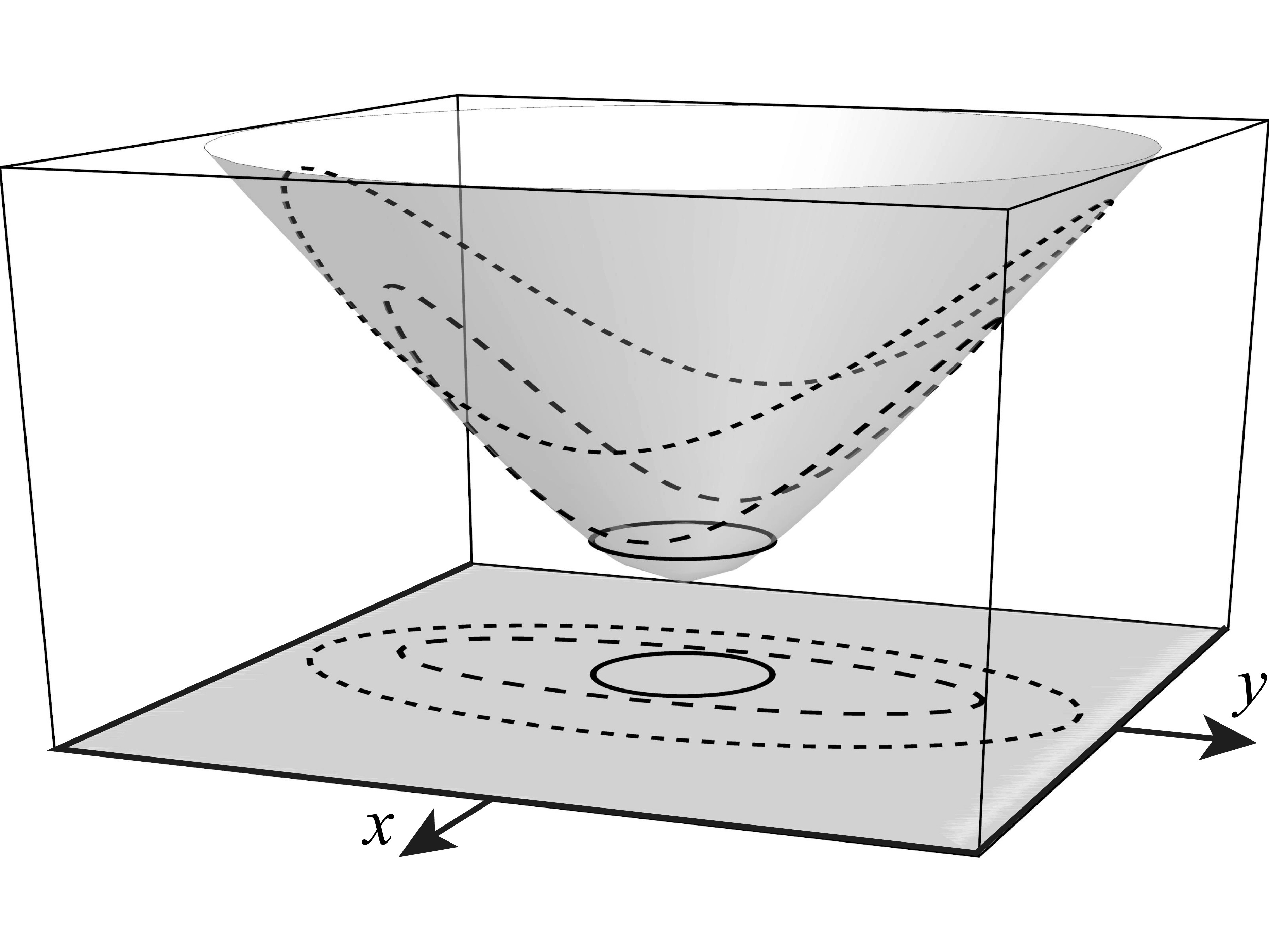}}
\caption[]{Trajectories $\tau(\varphi)$ on the lower half-hyperboloid
(of two sheets) given by \rf{HJ-42} with $\alpha=+1$ and $\beta=-2$,
and their projection on the full $x$--$y$ plane. The parameter values 
are all near to the cusp of the allowed region in Fig.\ \ref{fig:Epregion} (right):
$p_\phi=0.5$, $E=0.75$ (continuous line); $p_\phi=0.75$, $E=0.97$ (dashed line);
$p_\phi=0.9$, $E=0.993$ (dotted line).} 
\label{fig:trajectories_hyp2}
\end{figure}

Turning now to the case $\alpha>0$ for the pseudo-spherical system 
\rf{SystHp-I}, the separation of variables 
$S(\tau, \varphi) = S_1(\tau) +  p_\varphi\,\varphi$ yields
\be
      \frac{1+\alpha \varrho^2 \sinh^2\tau}{\varrho^2 \cosh^2\tau}  
      \left(\frac{\dd S_1}{\dd \tau}\right)^2 
        - \ii \beta \tanh\tau\left(\frac{\dd S_1}{\dd \tau}\right)
        +  \frac{p^2_\varphi}{\varrho^2 \sinh^2\tau}  = E, \lab{HJ-3h}
\ee    
so that the trajectory $\vartheta(\varphi)$ is found as
\be
        \sinh^2\tau = \frac{\alpha D}{1-\varepsilon \cos2\varphi}, \lab{HJ-42}
\ee 
lying on one sheet of a two-sheeted hyperboloid  $\varrho^2 = 1/\alpha$,
and where again $D$ and $\varepsilon$ are given in \rf{rrcuadrada}. The
orbits on this manifold are elliptic and are shown in 
Fig.\ \ref{fig:trajectories_hyp2}

\subsection{Separation in coordinate systems II and HII}
 
The second system of spherical coordinates $(\vartheta,\varphi)$
in \rf{Syst-II} leads to the Hamiltonian \rf{HJ-1} in the form
\bea
       &&\hskip-30pt\frac{1+\alpha R^2 \cos^2\vartheta}{R^2 \sin^2\vartheta} 
       	\left(\frac{\partial S}{\partial \vartheta}\right)^2 
        +\ii  \beta \cot\vartheta\left(\frac{\partial S}{\partial \vartheta}\right)
         \nonumber \\
		&&\hskip-20pt{}+\frac{1}{R^2 \sin^2\vartheta} 
			\left[ 1 + \frac{(1+\alpha R^2 )\cot^2\varphi}{\sin^2\vartheta} 
			\left(\frac{\partial S}{\partial \varphi}\right)^2 \right] 
			+ \ii \beta \frac{\cot\varphi}{\sin^2\vartheta}
				\left(\frac{\partial S}{\partial \varphi}\right)
     				 \nonumber \\
 		&&\hskip-20pt{}+ 2 \frac{1+\alpha R^2}{R^2 \sin^2\vartheta} \cot\vartheta  
 			\left(\frac{\partial S}{\partial \vartheta}\right) 
 			\cot\varphi\left(\frac{\partial S}{\partial \varphi}\right)    
        = E. \lab{HJ-5}
\eea  
When $\alpha<0$, separation of variables applies on the action function  
$S(\vartheta, \varphi) = S_1(\vartheta) + S_2(\varphi)$  and 
leads to the pair of equations
\bea
        &&\hskip-20pt -\alpha \left(\frac{\dd S_1}{\dd \vartheta}\right)^2 
                + \ii \beta  \cot\vartheta\left(\frac{\dd S_1}{\dd \vartheta}\right)
        +  \frac{K_\ssr{II}^2}{\sin^2\vartheta}  = E, \lab{HJ-6B}\\
        &&\hskip-20pt -\alpha \left(\frac{\dd S_2}{\dd \varphi}\right)^2 
        + \ii \beta \cot\varphi\left(\frac{\dd S_2}{\dd \varphi}\right)
                 = K_\ssr{II}^2, \lab{HJ-6A}  
\eea   
where $K_\ssr{II}^2$ is a separation constant. 
Rewriting  \rf{HJ-6A} in Cartesian $(x, y)$ coordinates, we obtain
\be
       \left[1 + \alpha (x^2+y^2)\right] \left(\frac{\dd S_2}{\dd y}\right)^2 
         - \ii \beta\, y \left(\frac{\dd S_2}{\dd y}\right)
        = K_\ssr{II}^2. \lab{HJ-7}
\ee  
The integration in $y$ yields a second 
integral of motion that depends on the parameters 
$(\alpha,\beta)$, 
\be
        I_2 := K_\ssr{II}^2 = \left[1 + \alpha (x^2+y^2)\right]  
        	p^2_y -  \ii \beta\,  y p_y. \lab{HJ-8}
\ee  

In the case $\alpha>0$, the action function admits separation 
of variables in the hyperbolic equidistant system HII in \rf{SystH-II},
$S(\tau_1, \tau_2) = S_1(\tau_1) + S_2(\tau_2)$ and yields the
two equations
\bea
        &&\hskip-20pt \alpha \left(\frac{\dd S_2}{\dd \tau_2}\right)^2 
        - \ii \beta \tanh\tau_2\left(\frac{\dd S_2}{\dd \tau_2}\right)
                 = K_\ssr{HII}^2, \lab{HJ-6Ah}\\
        &&\hskip-20pt \alpha \left(\frac{\dd S_1}{\dd \tau_1}\right)^2 
                - \ii \beta  \tanh\tau_1\left(\frac{\dd S_1}{\dd \tau_1}\right)
        +  \frac{K_\ssr{HII}^2}{\cosh^2\tau_1}  = E, \lab{HJ-6Bh}
\eea 
which lead to the same integral of motion $I_2$ in \rf{HJ-8}.


\subsection{Separation in the coordinate system III}
 
The third spherical system of coordinates in \rf{Syst-III} leads to 
the Hamilton-Jacobi form \rf{HJ-1} written as
\bea
     &&\hskip-30pt\frac{1+\alpha R^2 \cos^2\vartheta}{R^2 \sin^2\vartheta} 
     	\left(\frac{\partial S}{\partial \vartheta}\right)^2 
        + \ii \beta \cot\vartheta\left(\frac{\partial S}{\partial \vartheta}\right)
         \nonumber \\
	&&\hskip-20pt{}+\frac{1}{R^2 \sin^2\vartheta} 
		\left[ 1 + \frac{(1+\alpha R^2 )\tan^2\varphi}{\sin^2\vartheta} 
		\left(\frac{\partial S}{\partial \varphi}\right)^2 \right] 
		- \ii \beta \frac{\tan\varphi}{\sin^2\vartheta}
			\left(\frac{\partial S}{\partial \varphi}\right)
     	\nonumber \\
 	&&\hskip-20pt{}- 2 \frac{1+\alpha R^2}{R^2 \sin^2\vartheta} \cot\vartheta  
 		\left(\frac{\partial S}{\partial \vartheta}\right) 
 		\tan\varphi\left(\frac{\partial S}{\partial \varphi}\right)    
        = E. \lab{HJ-9}
\eea 
In the case $\alpha<0$, for $R^2 = -1/\alpha$, the separation 
of variables in the action function, $S(\vartheta, \varphi) 
= S_3(\vartheta) + S_4(\varphi)$ leads to
\bea
        &&\hskip-20pt -\alpha \left(\frac{\dd S_3}{\dd \vartheta}\right)^2 
                + \ii \beta \cot\vartheta\left(\frac{\dd S_3}{\dd \vartheta}\right)
                +  \frac{K_\ssr{III}^2}{\sin^2\vartheta}  = E, \lab{HJ-10B}\\
        &&\hskip-20pt -\alpha \left(\frac{\dd S_4}{\dd \varphi}\right)^2 
                - \ii \beta \tan\varphi\left(\frac{\dd S_4}{\dd \varphi}\right)
                 = K_\ssr{III}^2, \lab{HJ-10A}
\eea  
From \rf{HJ-10A} we find a third constant of motion 
that depends on $(\alpha,\beta)$, 
\be
        I_3 := K_\ssr{III}^2 = \left[1 + \alpha (x^2+y^2)\right]  
        	p^2_x  - \ii \beta x p_x, \lab{HJ-81}
\ee  
and which under the phase space $\onehalf\pi$-rotation 
$(x,p_x;y,p_y) \leftrightarrow (y,p_y;-x,-p_x)$ 
coincides with $I_2$ in \rf{HJ-8}. Finally, we note
that when $\alpha>0$, the separations of variables 
\rf{SystHp-I}--\rf{SystHp-III} on the hyperboloid yield the same 
integrals of motion $I_1$, $I_2$ and $I_3$ given above. 

We note that, unlike the three orthogonal coordinate systems
on the sphere, on hyperboloids there are {\it nine\/} 
orthogonal coordinate systems where the Laplace and the Helmholtz
equations yield to separation of variables \cite{Olevskii}.


\subsection{Separation of variables in the elliptic system} 

The Hamilton-Jacobi equation \rf{HJ-1} also yields
to separation in {\it elliptic\/} coordinates on the sphere 
in trigonometric  form \cite{PSW1,LS1,L1},

\be
        \xi^e_1 = R \cos\varphi\sqrt{1{-}k_1^2\cos^2\vartheta} ,\quad
        \xi^e_2 = R \sin\vartheta\sin\varphi, \quad
        \xi^e_3 = R \cos\vartheta\sqrt{1{-}k_3^2\cos^2\varphi} ,
                \label{COOR_ell}
\ee
where the constants $k_1:=\cos f$ and $k_3:=\sin f$ are related
to the interfocal distance $2f$ of the ellipses on the upper 
unit hemisphere, so that $k_1^2+k_3^2=1$.
When $\alpha<0$ and thus $R^2 = 1/|\alpha|$, 
the action function separates as $S(\vartheta,\varphi) = 
S^e_1(\vartheta)  + S^e_2(\varphi)$, and leads 
again to two equations,
\bea
        (1{-} k_1^2\cos^2\vartheta ) \left[ \alpha\, S_1^{e\,\prime\,2}
                + \ii \beta \tan\vartheta \, S_1^{e\,\prime}\right]
                + E k_1^2\sin^2\vartheta &=& K_e^2,\\
        \alpha (1 {-} k_3^2\cos^2\varphi) S_2^{e\,\prime\,2}
                - \ii \beta  k_3^2 \cos\varphi \sin \varphi\,S_2^{e\,\prime}
                        + E k_3^2\sin^2\varphi &=& -K_e^2,{\quad}
\eea
where $K_e^2$ is a separation constant, $S_1^{e\,\prime}:=\dd S_1/\dd\vartheta$ 
and $S_2^{e\,\prime}:=\dd S_2/\dd\varphi$.
Eliminating $E$ from these equations one obtains
\bea
	&&{\hskip-30pt} K_e^2(k_3^2\sin^2\varphi + k_1^2\sin^2\vartheta)\nonumber\\
	&=&k_3^2\sin^2\varphi (1-k_1^2\cos^2\vartheta)
                \left[\alpha \left(S'_1\right)^2 
                + \ii \beta  \tan\vartheta S'_1 \right]	\lab{elloc}\\
		&&{}- k_1^2\sin^2\vartheta\left[ \alpha (1-k_3^2\cos^2\varphi) 
                \left(S'_2\right)^2 -  \ii \beta k_3^2 \cos\varphi 
                	\sin \varphi S'_2 \right].\nonumber
\eea
Returning to Cartesian $(x,y)$ coordinates,
\bea
	\frac{S_1^{e\,\prime}}{R} &=& p_x 
		\frac{k_1^2\sin\vartheta\cos\vartheta\cos\varphi
			}{\sqrt{1-k_1^2\cos^2\vartheta}}
		+ p_y\cos\vartheta\sin\varphi, \lab{S1}\\
	\frac{S_2^{e\,\prime}}{R} &=& -p_x 
		\sqrt{1-k_1^2\cos^2\vartheta}\,\sin\varphi
		+p_y\sin\vartheta\cos\varphi, \lab{S2}
\eea
we can express the constant $K_e^2$ as
\be
   \begin{array}{rcl}
        K_e^2 &=& -\alpha k_1^2 y^2 p_x^2 + 2\alpha k_1^2 xy p_x p_y\\[3pt] 
        	&&\displaystyle {\!} -  \Big(\alpha k_1^2 x^2 + k_3^2 
        		[1 +\alpha(x^2+y^2)]\Big) p^2_y 
                +  \ii \beta k_3^2 y p_y   \\[3pt] 
        &=& -\alpha k_1^2 I_1^2 - k_3^2 I_2. \end{array} \lab{C2expr}
\ee
Thus the elliptic separation constant $K_e^2$ is 
not functionally independent but depends on the 
constants $I_1$ and $I_2$ in \rf{I1} and \rf{HJ-8}.


\section{Algebraic structure and conclusions}  \label{sec:five}

We have found three functionally independent integrals of motion, 
$I_1$ in \rf{I1}, $I_2$ in \rf{HJ-8}, and $I_3$ in \rf{HJ-81} with
 no singularities on the full $(\alpha,\beta)$
parameter space. To probe their algebraic structure let us define   
\bea
        J_1 &:=& \onehalf I_1 = \onehalf (x p_y - y p_x), \lab{HJ-91}\\
        J_2 &:=& \onehalf (I_3 - I_2) 
                = \onehalf \left[1 + \alpha (x^2+y^2)\right] (p^2_x - p_y^2) 
                - \onehalf \ii \beta ( x p_x - y p_y).{\qquad}
						\lab{HJ-911} 
\eea
The function $J_1$ is $\onehalf$-angular momentum and its
Poisson operator $\{J_1,\circ\}$ generates rotations of 
phase space, while the function $J_2$ does depend 
on $(\alpha,\beta)$.  
These functions Poisson-commute with the Zernike 
Hamiltonian function $H^{(\alpha,\beta)}$ in \rf{H1}, 
which can be written as
\be 
        H^{(\alpha,\beta)} =   I_3 + I_2 - \alpha I_1^2 , \lab{HJ-10}
\ee
but do not commute with each other. This shows that the 
generalized classical $(\alpha,\beta)$-Hamiltonian 
of Zernike, $H^{(\alpha,\beta)}$ in \rf{H3}, 
is superintegrable on each of the domains examined
above, in particular on the $(x,y)$-disk 
${\cal D}_\ssr{R}$, $r<R=1/\surd|\alpha|$ for 
$\alpha<0$, that contains the Zernike original case
$(\alpha_\ssr{Z},\beta_\ssr{Z})=(-1,-2)$.

To identify the symmetry of the generalized Zernike
$(\alpha,\beta)$-Hamiltonians, we introduce a new 
integral of motion through the Poisson bracket of 
\rf{HJ-91} and \rf{HJ-911},  
\be
       J_3:= \{J_1, J_2\} =\left[1 + \alpha (x^2+y^2)\right] p_x p_y 
       		- \ii \onehalf\beta(x p_y + y p_x),  
                \lab{HJ-11} 
\ee
which also Poisson-commutes with $H^{(\alpha,\beta)}$,
and is functionally independent of $J_1$ and $J_2$,
although it can be seen that $J_2$ and $J_3$ are 
connected to each other by a rotation of 
$\frac14\pi$ in the $x$--$y$ phase space planes. 
The algebraic structure of three functions 
$J_1,\, J_2,\, J_3$ is thus found to be
\bea
        \{J_3, J_1\} &=& J_2, \qquad \{J_1, J_2\}=J_3, \lab{HJ-13}\\
        \{J_2, J_3\} &=& J_1(\beta^2 - 
        2\alpha H^{(\alpha,\beta)} - 8\alpha^2 J_1^2). 
        	\lab{HJ-131}
\eea  
They  form therefore a cubic Higgs algebra \cite{Higgs} that 
Poisson-commutes with the generalized Zernike 
Hamiltonian, $\{J_i,H^{(\alpha,\beta)}\}=0$. 

	When $\alpha\to0$ so $R\to\infty$, the Zernike 
Hamiltonian becomes a simpler quadratic function,
\be 
	H^{(0,\beta)}({\bf q},{\bf p})
		={\bf p}^2-\ii\beta\,{\bf q}\cdot{\bf p}.
		\lab{Hin0} 
\ee
The Poisson operators of all quadratic functions of
these four phase space coordinates close under
commutation into the real symplectic Lie algebra {\sf sp($4$,R)}.

The Hamiltonian \rf{Hin0} belongs to the elliptic orbit 
of harmonic oscillators \cite[Chap.\ 12]{GeomOpt},
as can be seen under the complex linear canonical transformation
\be 
	\vecdos{{\bf p}}{{\bf q}}
		= \matdos{{\bf1}/\surd2
		}{\ii\beta{\bf1}/\surd2}{{\bf0}}{\surd2\,{\bf1}}
		\vecdos{{\bf P}}{{\bf Q}}.
			\lab{cantfm}
\ee
This maps \rf{Hin0} on a regular harmonic oscillator,
\be 
	F_0:=H^{(0,\beta)}({\bf Q},{\bf P}) 
		= \onehalf({\bf P}^2+\beta^2{\bf Q}^2),
			\lab{regHO}
\ee
and the three constants of the motion, $J_1,\,J_2,\,J_3$ in \rf{HJ-91}, \rf{HJ-911} and 
\rf{HJ-11}, on
\bea 
	F_1&:=&\onehalf\,{\bf Q}\times{\bf P}=\onehalf(Q_xP_y-Q_yP_x), \lab{F1}\\
	F_2&:=&\onehalf(P_x^2+\beta Q_x^2) - \onehalf (P_y^2+\beta^2Q_y^2), \lab{F2}\\
	F_3&:=&\onehalf P_xP_y+\onehalf\beta^2\,Q_xQ_y \lab{F3}
\eea
whose Poisson brackets close into a scaled {\sf u($2$)} Lie algebra,
\be 
	\{F_1,F_2\}=F_3,\quad \{F_2,F_3\}=\beta^2F_1,\quad \{F_3,F_1\}=F_2,
		\quad\ \{F_0,F_i\}=0.   \lab{Fourier-U2}
\ee

In the paraxial geometric or wave optical interpretation, the central 
$F_0\in{\sf u({\rm1})}$ generates isotropic fractional
Fourier transforms \cite{Simon-KBW}, while $F_2$ generates anisotropic
ones, $F_1$ generates rotations, and $F_3$ generates gyrations 
\cite{Alieva-Bastiaans} that transform Hermite-Gauss into Laguerre-Gauss
beams. Together their Poisson operators form the {\it Fourier\/} algebra
\cite{Simon-KBW}, which is the maximal compact subalgebra in {\sf sp($4$,R)}.
If $\beta$ were a pure imaginary number, \rf{regHO} would be the repulsive
oscillator Hamiltonian and \rf{F1}--\rf{F3} its commuting `Fourier' algebra 
${\sf su({\rm1,1})}={\sf so({\rm2,1})}$; a similar treatment of the classical 
system with Hamiltonian \rf{H1} would yield hyperbolic orbits. For $\beta=0$ 
a free system with an inhomogeneous {\sf iso({\rm2})} `Fourier' algebra
would appear.

The original Zernike system $\hat Z^{(\alpha_\ssr{Z},\beta_\ssr{Z})}$ 
in \rf{Zernikeq} \cite{Zernike34} was proposed to develop a set 
orthogonal and complete set of two-variable orthogonal polynomials 
$Z_{n,m}(r)\exp(\ii m\phi)$, $Z_{n,m}(1)=1$, $|m|\le n$, which 
present the same $(n,m)$-pattern as the two-dimensional quantum harmonic
oscillator states. There has been some effort in replicating the raising
and lowering techniques of the oscillator scheme on the Zernike system 
\cite{Wunsche,Shakibaei} without achieving a proper Lie algebra. Because
here we have a two-parameter system $H^{(\alpha,\beta)}$, we could surmise
that superintegrable systems can be obtained as a new kind of algebra
deformation, from \rf{regHO}--\rf{Fourier-U2} to \rf{HJ-91}--\rf{HJ-131},
consisting in the addition of the square of an element of
a Lie algebra to the generator designed to be the original quadratic 
Hamiltonian. Imposing boundary conditions such as those proposed by 
Zernike will need the quantum treatment of this construction.

\section*{Acknowledgements}

We acknowledge the interest and early discussions with Prof.\ Natig
M.\ Atakishiyev (Instituto de Matem\'aticas, {\sc unam});
we thank Guillermo Kr\"otzsch ({\sc icf-unam}) for indispensable help
with the figures. G.S.P.\ and A.Y.\ thank the support of project 
{\sc pro-sni-2016} (Universidad de Guadalajara). 
K.B.W.\ thanks Cristina Salto-Alegre (Posgrado
en Ciencias F\'isicas, {\sc icf-unam}) for her interest and interaction
on the subject, and acknowledges the support of {\sc unam-dgapa} 
Project {\it \'Optica Matem\'atica\/} {\sc papiit}-IN101115.


\end{document}